\documentclass{amsart}
\usepackage{graphicx}
\theoremstyle{definition}

\theoremstyle{remark}

\numberwithin{equation}{section}

\begin{document}

\title{On the Scaling Window of Model RB}%
\author{Chunyan Zhao$^1$}%
\address{$^{1}$School of Science, Beijing University of Aeronautics and Astronautics, Beijing 100083, China\\
Key Laboratory of Mathematics, Informatics and Behavioral
Semantics, Ministry of Education}%
\email{xiaoyanzi@ss.buaa.edu.cn}%
\author{Ke Xu$^{2}$}%
\address{$^{2}$School of Computers, Beijing University of Aeronautics and Astronautics, Beijing 100083, China\\
National Laboratory of Software Development Environment}%
\email{kexu@nlsde.buaa.edu.cn}%
\author{Zhiming Zheng$^{3}$}%
\address{$^{3}$School of Science, Beijing University of Aeronautics and Astronautics, Beijing 100083, China\\
Key Laboratory of Mathematics, Informatics and Behavioral
Semantics, Ministry of Education}%
\email{zzheng@pku.edu.cn}
\thanks{Supported in part by National 973 Program of China(Grant No. 200532CB1902)
and NSFC(Grant Nos. 60473109 and 60403003).}%

\begin{abstract}
This paper analyzes the scaling window of a random CSP model (i.e.
model RB) for which we can identify the threshold points exactly,
denoted by $r_{cr}$ or $p_{cr}$. For this model, we establish the
scaling window $W(n,\delta)=(r_{-}(n,\delta),\ \ r_{+}(n,\delta))$
such that the probability of a random instance being satisfiable
is greater than $1-\delta$ for $r<r_{-}(n,\delta)$ and is less
than $\delta$ for $r>r_{+}(n,\delta)$. Specifically, we obtain the
following result
$$W(n,\delta)=(r_{cr}-\Theta(\frac{1}{n^{1-\varepsilon}\ln n}), \ \
r_{cr}+\Theta(\frac{1}{n\ln n})),$$ where $0\leq\varepsilon<1$ is
a constant. A similar result with respect to the other parameter
$p$ is also obtained. Since the instances generated by model RB
have been shown to be hard at the threshold, this is the first
attempt, as far as we know, to analyze the scaling window of such
a model with hard instances.\\

\end{abstract}
\maketitle

\section{Introduction}
The Constraint Satisfaction Problem(CSP), originated from
artificial intelligence, has become an important and active field
of statistical physics, information theory and computer science.
The CSP area is very interdisciplinary, since it embeds ideas from
many research fields, like artificial intelligence, databases,
programming languages and operation research. A constraint
satisfaction problem consists of a finite set
$U=\{u_{1},u_{2},\cdots,u_{n}\}$ of $n$ variables, each $u_{i}$
associated with a domain of values $D_{i}$, and a set of
constraints. Each of the constraints $C_{i_{1}i_{2}\cdots i_{k}}$
is a relation, defined on some subset $\{u_{i_{1}},
u_{i_{2}},\cdots, u_{i_{k}} \}$ of $n$ variables, called its
scope, denoting their legal tuples of values. A \emph{solution} to
a CSP is an assignment of a value to each variable from its domain
such that all the constraints of this CSP are satisfied. A
constraint is said to be \emph{satisfied} if the tuple of values
assigned to the variables in this constraint is a legal one. A CSP
is called \emph{satisfiable} if and only if it has at least one
solution. The task of a CSP is to find a solution or to prove that no solution exists.\\
\indent Given a CSP, we are interested in polynomial-time
algorithms, that is, algorithms whose running time is bounded by a
polynomial in the number of variables. Cook's Theorem\cite{B}
asserts that satisfiability is NP-complete and at least as hard as
any problem whose solutions can be verified in polynomial time.
Most of the interesting CSPs are NP-complete problems. We know
that $k$-SAT problem is a canonical version of the CSPs, in which
variables can be assigned the value True or False(called Boolean
variables). A lot of efforts have been devoted to $k$-SAT and it
is widely believed that no efficient algorithm exists for $k$-SAT.
However, it is shown that most instances of $k$-SAT can be solved
efficiently, so perhaps genuine hardness is only present in a tiny
fraction of all instances. In 1990s, a remarkable
progress\cite{Che, S, K, M} was made that the the really difficult
instances is related to phase transition phenomenon, as suggested
in the pioneering work of Fu and Anderson\cite{F}. The study of
phase transitions has attracted much interest subsequently\cite{Ho,M}.\\
\indent In recent years, random $k$-SAT has been well studied both
from theoretical and algorithmic point of views. If $k=2$ then it
is known that there is a satisfiability threshold at
$\alpha_{c}=1$ (here $\alpha$ represents the ratio of clauses $m$
to variables $n$), below which the probability of a random
instance being satisfiable tends to $1$ and above which it tends
to $0$ as $n$ approaches infinity\cite{Chv}. This was sharpened
in\cite{Go, V}. Random $2$-SAT is now pretty much understood.
However, for $k\geq 3$, the existence of the phase transition
phenomenon has not been established, not even the exact value of
the threshold point\cite{A,Ki}. \\
\indent To gain a better understanding of how the phase transition
scales with problem size, the finite-size scaling method has been
introduced from statistical mechanics\cite{K, Ge}. We use
finite-size scaling, a method from statistical physics in which
observing how the width of a transition narrows with increasing
sample size gives direct evidence for critical behavior at a phase
transition. Finite-size scaling is the study of changes in the
transition behavior due to finite-size effects, in particular,
broadening of the transition region for finite $n$. More
precisely, for $0<\delta<1$, let $r_{-}(n,\delta)$ be the supremum
over $r$ such that the probability of a random CSP instance being
satisfiable is at least $1-\delta$, and similarly, let
$r_{+}(n,\delta)$ be the infimum over $r$ such that the
probability of a random CSP instance being satisfiable is at most
$\delta$. Then, for $r$ within the scaling window
$W(n,\delta)=(r_{-}(n,\delta),\ \ r_{+}(n,\delta))$ the
probability is between $\delta$ and $1-\delta$. And for all
$\delta$, $|r_{+}(n,\delta)-r_{-}(n,\delta)|\rightarrow0$ as
$n\rightarrow\infty$. For random 2-SAT, it has been determined
that the scaling window is $W(n,\delta)=(1-\Theta(n^{-1/3}),\ \
1+\Theta(n^{-1/3}))$\cite{B}.\\
\indent Model RB is a random CSP model proposed by Xu and Li to
overcome the trivial insolubility of standard CSP
models\cite{XWe}. For this model, we can not only establish the
existence of phase transitions, but also pinpoint the threshold
points exactly, denoted by $r_{cr}$ or  $p_{cr}$. Moreover, it has
been proved that almost all instances of model RB have no
tree-like resolution proofs of less than exponential size
\cite{XWe}. This implies that unlike random 2-SAT, model RB can be
used to generate hard instances, which has also been confirmed by
experiments\cite{Ge}. Motivated by the work on the scaling window
of random 2-SAT, in this paper, we study the scaling window of
model RB and obtain that
$W(n,\delta)=(r_{cr}-\Theta(\frac{1}{n^{1-\varepsilon}\ln n}),\ \
r_{cr}+\Theta(\frac{1}{n\ln n}))$. And we also obtain
similar results about the other control parameter $p$.\\
\indent The main contribution of this paper is not to present 
new methods for computing the scaling window, but to show that 
for an interesting model with hard instances (i.e. model RB), 
not only can the threshold points be located exactly, but also 
the scaling window can be deteremined using standard methods. 
This means that hopefully, more mathematical properties about 
the threshold behavior of model RB can be obtained in a relatively 
easy way, which will help to shed light on the phase transition 
phenomenon in NP-complete problems. 
The rest of the paper is organized as follows. In the next
section, we will give a brief introduction about model RB. The
main results of this paper and their proofs will be given in
Section 3 and Section 4 respectively. Finally, we will conclude in
Section 5.
\section{Model RB}
\indent We can pinpoint the threshold location for model RB
proposed by Xu and Li\cite{XWe}. The way of generating random
instances for model RB is:\\
\\
\emph{(1). Given a set $U$ of $n$ variables, select with
repetition $m=rn\ln n$ random constraints. Each random constraint
is formed by selecting without repetition $k$ of $n$ variables,
where $k\geq
2$ is an integer.\\
(2). Next, for each constraint we select uniformly at random
without repetition $q=p\cdot d^{k}$ illegal tuples of values,
i.e., each constraint contains exactly $(1-p)\cdot d^{k}$ legal
ones, where $d=n^{\alpha}$ is the domain size of each variable and
$\alpha>0$ is a constant. \\}
\\
\indent In this paper, the probability of a random CSP instance
being satisfiable is denoted by Pr(Sat). It is proved that for
model RB the phase transition phenomenon occurs at
$r_{cr}=-\frac{\alpha}{\ln(1-p)}$ or
$p_{cr}=1-e^{-\frac{\alpha}{r}}$ as $n$ approaches
infinity\cite{XWe}. More precisely, we have the following two theorems.\\
\\
\emph{\textbf{Theorem 2.1}\cite{XWe} Let
$r_{cr}=-\frac{\alpha}{\ln (1-p)}$. If $\alpha>\frac{1}{k}$,
$0<p<1$ are two constants and $k$, $p$ satisfy the inequality
$k\geq\frac{1}{1-p}$, then
$$\lim_{n\rightarrow\infty}Pr(Sat)=1\ \ when\ \ r<r_{cr},$$
$$\lim_{n\rightarrow\infty}Pr(Sat)=0\ \ when\ \ r>r_{cr}.$$
\textbf{Theorem 2.2}\cite{XWe} Let
$p_{cr}=1-e^{-\frac{\alpha}{r}}$. If $\alpha>\frac{1}{k}$, $r>0$
are two constants and $k$, $\alpha$ satisfy the inequality
$ke^{-\frac{\alpha}{r}}\geq1$, then
$$\lim_{n\rightarrow\infty}Pr(Sat)=1\ \ when\ \  p<p_{cr},$$
$$\lim_{n\rightarrow\infty}Pr(Sat)=0\ \ when\ \ p>p_{cr}.$$}
\section{Main results}
\indent Our main results are the following two theorems.\\
\\
\textbf{Theorem 3.1} For all sufficiently small $\delta>0$, there
exist $r_{-}(n,\delta)$ and $r_{+}(n,\delta)$ such that the
following holds:\\
$$Pr(Sat)>1-\delta,\ \ when\ \  r<r_{-}(n,\delta) ;$$
$$Pr(Sat)\ <\ \delta,\ \ \ \ \ \ when\ \  r>r_{+}(n,\delta) ,$$
where $r_{-}(n,\delta)=r_{cr}-\Theta(\frac{1}{n^{1-\varepsilon}\ln
n})$, $r_{+}(n,\delta)=r_{cr}+\Theta(\frac{1}{n\ln n})$. So that
the scaling window of model RB is\\
\begin{equation*}
W(n,\delta)=(r_{cr}-\Theta(\frac{1}{n^{1-\varepsilon}\ln n}),\ \
r_{cr}+\Theta(\frac{1}{n\ln n})).
\end{equation*}
It is easy to see that
$|r_{+}(n,\delta)-r_{-}(n,\delta)|\rightarrow0$, as
$n\rightarrow\infty$.\\
\\
\textbf{Theorem 3.2} For all sufficiently small $\delta>0$, there
exist $p_{-}(n,\delta)$ and $p_{+}(n,\delta)$ such that the
following holds:\\
$$Pr(Sat)>1-\delta,\ \ when\ \  p<p_{-}(n,\delta) ;$$
$$Pr(Sat)\ <\ \delta,\ \ \ \ \ \ when\ \  p>p_{+}(n,\delta) ,$$
where $p_{-}(n,\delta)=p_{cr}-\Theta(\frac{1}{n^{1-\varepsilon}\ln
n})$, $p_{+}(n,\delta)=p_{cr}+\Theta(\frac{1}{n\ln n})$. So that
the
scaling window of Model RB is\\
\begin{equation*}
W(n,\delta)=(p_{cr}-\Theta(\frac{1}{n^{1-\varepsilon}\ln n}),\ \
p_{cr}+\Theta(\frac{1}{n\ln n})).
\end{equation*}
It is not difficult to see that
$|p_{+}(n,\delta)-p_{-}(n,\delta)|\rightarrow0$, as
$n\rightarrow\infty$.\\
\\
\textbf{Remark 3.1} If $n\rightarrow\infty$, then
$r_{+}(n,\delta),r_{-}(n,\delta)\rightarrow r_{cr}$,
$p_{+}(n,\delta),p_{-}(n,\delta)\rightarrow p_{cr}$. For every
sufficiently small $\delta$, Theorem 3.1 and Theorem 3.2 hold. So
we can obtain
$$\lim_{n\rightarrow\infty}Pr(Sat)=1\ \ when\ \ r<r_{cr}\ \ or\ \ p<p_{cr},$$
$$\lim_{n\rightarrow\infty}Pr(Sat)=0\ \ when\ \ r>r_{cr}\ \ or\ \ p>p_{cr}.$$
This is the result of Xu and Li\cite{XWe}.
\section{Proof of the results}
\indent To prove the main results, we need the following lemmas.\\
\\
\textbf{Lemma 4.1} Let $c=\alpha+1-r_{cr}kp$, then $c<1$. \\
\\
\textbf{Proof} We know that $r_{cr}=-\frac{\alpha}{\ln(1-p)}$,
then
\begin{eqnarray}
c&=&\alpha+1+\frac{\alpha kp}{\ln(1-p)}\nonumber\\
 &=&1+\frac{\alpha[kp+\ln(1-p)]}{\ln(1-p)}\nonumber
\end{eqnarray}
\indent Assume that $f(p)=kp+\ln(1-p)$, hence we have
$f'(p)=-\frac{1}{1-p}+k$.\\
\indent By the condition of Theorem 2.1, we have
$k\geq\frac{1}{1-p}$, hence $f'(p)\geq0$. That is $f(p)$ is a
monotone increasing function.\\
\indent So $f(p)>f(0)$, that is $kp+\ln(1-p)>0$. It is obvious
that $\ln(1-p)<0$ because of $0<p<1$. And
$\alpha>\frac{1}{k}$ is a constant.\\
\indent Hence $\frac{\alpha[kp+\ln(1-p)]}{\ln(1-p)}<0$.\\
\indent Therefore, it is proved that
$c=1+\frac{\alpha[kp+\ln(1-p)]}{\ln(1-p)}<1$.\\
\\
\textbf{Lemma 4.2} Let $c=\alpha+1-rkp_{cr}$, then $c<1$.\\
\\
\textbf{Proof} We know that $p_{cr}=1-e^{-\frac{\alpha}{r}}$, so \\
\begin{eqnarray}
c&=&\alpha+1-rk(1-e^{-\frac{\alpha}{r}})\nonumber\\
&=&1-r[-\frac{\alpha}{r}+k(1-e^{-\frac{\alpha}{r}})]\nonumber
\end{eqnarray}
\indent Let $-\frac{\alpha}{r}=x$, then $x\in(-\infty,0)$.
Suppose $h(x)=x+k(1-e^{x})$, then $h'(x)=1-ke^{x}$.\\
\indent By the condition of Theorem 2.2,
$ke^{x}=ke^{-\frac{\alpha}{r}}\geq1$, hence $h'(x)\leq0$. That is $h(x)$ is a monotone decreasing function.\\
\indent So $h(x)>h(0)$, that is $h(x)>0$. And $r>0$ is a
constant, hence it is proved that $c=1-r[-\frac{\alpha}{r}+k(1-e^{-\frac{\alpha}{r}})]<1$.\\
\\
\textbf{Proof of Theorem 3.1} Let $N$ denote the number of
satisfying
assignments for a random CSP instance, we can obtain that\\
\begin{eqnarray}
E(N)&=& d^{n}(1-p)^{rn\ln n}\nonumber\\
    &=& n^{\alpha n}(1-p)^{rn\ln n}\
\end{eqnarray}
\indent Assume that $E(N)<\delta$, by (1) we get
\begin{equation}
[\alpha+r\ln(1-p)]n\ln n<\ln\delta
\end{equation}
\begin{equation}
\alpha+r\ln(1-p)<\frac{\ln\delta}{n\ln n}
\end{equation}
\begin{equation}
r>-\frac{\alpha}{\ln(1-p)}+\frac{\ln\delta}{n\ln
n\ln(1-p)}=r_{cr}+\frac{\ln\delta}{n\ln n\ln(1-p)}
\end{equation}
Using the Markov inequality Pr(Sat)$\leq E(N)$, we get
Pr(Sat)$<\delta$ for
\begin{equation}
r>r_{cr}+\Theta(\frac{1}{n\ln n}).\ \
\end{equation}
\indent Here note that $f=\Theta(g)$ represents there exist two
finite constants $c_{1}>0$ and $c_{2}>0$ such that
$c_{1}<f/g<c_{2}$.\\
\indent In the following, we use Cauchy
inequality Pr(Sat)$\geq \frac{E^{2}(N)}{E(N^{2})}$ to prove when
$r<r_{cr}+\Theta(\frac{1}{n\ln n})$,
we have Pr(Sat)$>1-\delta$.\\
\indent In the remaining part of the paper, the expression of
$E(N^2)$ will play an important role in the proof of the main
results. The derivation of this expression can be found in
\cite{XWe}. For the convenience of the reader, we give an outline
of it as follows.\\
\textbf{Definition 4.1} Let $\langle t_{i},t_{j}\rangle$
represents an ordered assignment pair to the $n$ variables in $U$,
which satisfies a CSP instance if and only if both $t_{i}$ and
$t_{j}$ satisfy the CSP instance. And $P(\langle
t_{i},t_{j}\rangle)$ denotes the probability of $\langle
t_{i},t_{j}\rangle$ satisfying
a CSP instance.\\
\textbf{Definition 4.2} The similarity number $S$ of an assignment
pair $\langle t_{i},t_{j}\rangle$ is the number of variables
$t_{i}$ and $t_{j}$ take the identical values. It is obvious that
$0\leq S\leq n$, and let $s=\frac{S}{n}$. Let $A_{S}$ be the set
of
assignments whose similarity number is equal to $S$.\\
\indent We can get the expression of $E(N^{2})$ is
\begin{eqnarray}
|A_{S}|P(\langle t_{i},t_{j}\rangle) &=&
\sum_{S=0}^{n}|A_{S}|P(\langle t_{i},t_{j}\rangle)\nonumber\\
&=& d^{n}{n \choose S}(d-1)^{n-S}[\frac{{d^{k}-1 \choose
q}}{{d^{k} \choose q}}\cdot \frac{{S \choose k}}{{n \choose
k}}+\frac{{d^{k}-2 \choose q}}{{d^{k} \choose q}}\cdot (1-\frac{{S
\choose k}}{{n \choose k}})]^{rn\ln n}\nonumber
\end{eqnarray}
\indent First we need to estimate $E(N^{2})$. We can rewrite the
above equation as the following one
\begin{equation}
\begin{array}{ll}
|A_{S}|P(\langle t_{i},t_{j}\rangle)
=&E^{2}(N)[1+\frac{p}{1-p}(s^{k}+\frac{g(s)}{n})]^{rn\ln n}\cdot\\
  &(1-\frac{1}{n^{\alpha}})^{n-ns}(\frac{1}{n^{\alpha}})^{ns}{n \choose
ns}(1+O(\frac{1}{n}))
\end{array}
\end{equation}
where $g(s)=\frac{k(k-1)(s^{k}-s^{k-1})}{2}$.\\
\indent When $n$ is sufficiently large, except $E^{2}(N)$, the
dominant contribution to (4.6) comes from
\begin{eqnarray}
f(s)&=&
(1+\frac{p}{1-p}s^{k})^{rn\ln n}(\frac{1}{n^{\alpha}})^{ns}\nonumber\\
    &=& e^{[r\ln(1+\frac{p}{1-p}s^{k})-\alpha s]n\ln n}
\end{eqnarray}
\indent We put $h(s)=r\ln(1+\frac{p}{1-p}s^{k})-\alpha s$ and
focus on the function $h(s)$, differentiating $h(s)$ twice with
respect to $s$ we get
\begin{equation}
h''(s)=\frac{rkps^{k-2}[(k-1)(1-p)-ps^{k}]}{(1-p+ps^{k})^{2}}
\end{equation}
\indent Applying the condition $k\geq\frac{1}{1-p}$, we get
$(k-1)(1-p)-ps^{k}\geq0$ on the interval $[0,1]$, then
$h''(s)\geq0$. So $h(s)$ is a convex function. It is easy to see
that $h(0)=0$ and $h(1)=-r\ln(1-p)-\alpha$. So when
$r<r_{cr}-\Theta(1/(n^{1-\varepsilon}\ln n))$, we have
$h(1)\leq0$. On the interval $0<s<1$, we get $h(s)<0$. So there
exist $0<\delta_{1}<1$ and $0<\delta_{2}<1$ such that when
$r<r_{cr}-\Theta(1/(n^{1-\varepsilon}\ln n))$, $h(s)$ is mainly
decided by the values $s\in[0,\delta_{1}]\cup[1-\delta_{2},1]$. So
we only need to consider those terms
$s\in[0,\delta_{1}]\cup[1-\delta_{2},1]$ to estimate (4.6). This
is different from the proof in Xu and Li\cite{XWe} for
establishing the existence of phase transitions, where only those
terms
$s\in[0,\delta_{1}]$ were considered.\\
\\
\indent (i) $s\in[0,\delta_{1}]$\\
\\
\indent We can learn from Xu and Li\cite{XWe} that\\
\begin{equation}
\sum_{s\in[0,\delta_{1}]}|A_{S}|P(\langle t_{i},t_{j}\rangle)\leq
E^{2}(N)(1+O(\frac{1}{n}))
\end{equation}
\indent (ii) $s\in[1-\delta_{2},1]$\\
\\
\indent It is easily known that if $s\in[1-\delta_{2},1]$, we can
obtain $s^{k}-s^{k-1}<0$, thus
$g(s)=\frac{k(k-1)(s^{k}-s^{k-1})}{2}<0$. So we can get the
following inequality
\begin{eqnarray}
|A_{S}|P(\langle t_{i},t_{j}\rangle) & \leq &
E^{2}(N)(1+\frac{p}{1-p}s^{k})^{rn\ln
n}\nonumber\\
 & & \cdot(1-\frac{1}{n^{\alpha}})^{(n-ns)}(\frac{1}{n^{\alpha}})^{ns}{n\choose
ns}(1+O(\frac{1}{n}))\nonumber\\
                                     & = &  E(N)(1-p+ps^{k})^{rn\ln
n}(n^{\alpha}-1)^{n-ns}{n\choose ns}(1+O(\frac{1}{n}))
\end{eqnarray}
\indent When $s=1(S=n)$, we obtain
\begin{equation}
|A_{S}|P(\langle t_{i},t_{j}\rangle)=E(N)(1+O(\frac{1}{n}));
\end{equation}
\indent When $s=\frac{n-t}{n}(S=n-t)$, where $1\leq t\ll n$. We
can get that
\begin{eqnarray}
|A_{S}|P(\langle t_{i},t_{j}\rangle) &\leq&
E(N)\cdot[1-p+p(\frac{n-t}{n})^{k}](n^{\alpha}-1)^{t}{n \choose t}\cdot(1+O(\frac{1}{n}))\nonumber\\
&\leq& E(N)e^{-p[1-(\frac{n-t}{n})^{k}]^{rn\ln
n}}(n^{\alpha}-1)^{t}{n \choose t}\cdot(1+O(\frac{1}{n}))\nonumber\\
&\leq&
E(N)\frac{n^{(\alpha+1)t}}{n^{rnpt(\frac{k}{n}-O(\frac{1}{n^{2}}))}}(1+O(\frac{1}{n}))\nonumber\\
&=& E(N)\frac{n^{(\alpha+1)t}}{n^{rkpt-O(\frac{1}{n})}}(1+O(\frac{1}{n}))\nonumber\\
&\leq&
E(N)(\frac{n^{\alpha+1+O(\frac{1}{n})}}{n^{rkp}})^t(1+O(\frac{1}{n}))
\end{eqnarray}
\indent When $n$ is sufficiently large, let
$c=\alpha+1-r_{cr}kp=\alpha+1+\frac{\alpha kp}{\ln(1-p)}$.
Thus it is divided into two cases to discuss the value of $c$.\\
\textbf{Case 1:} $c<0$. \\
\indent When $s=\frac{n-1}{n}$, by (4.12) we can obtain
\begin{equation}
|A_{S}|P(\langle t_{i},t_{j}\rangle)\leq E(N)\cdot
n^c\cdot(1+O(\frac{1}{n}))
\end{equation}
\indent When $s=\frac{n-2}{n}$, by (4.12) we have
\begin{equation}
|A_{S}|P(\langle t_{i},t_{j}\rangle)\leq E(N)\cdot
n^{2c}\cdot(1+O(\frac{1}{n}))
\end{equation}
$\cdots  \  \  \cdots   \  \  \cdots$\\
\indent So we can get
\begin{eqnarray}
\sum_{s\in[1-\delta_{2},1]}|A_{S}|P(\langle t_{i},t_{j}\rangle)
&\leq& E(N)(1+n^2+n^{2c}+\cdots)\cdot(1+O(\frac{1}{n}))\nonumber\\
&=& E(N)(1+O(n^c))
\end{eqnarray}
\indent It is shown from (i) and (ii) that
\begin{eqnarray}
E(N^{2}) &=& \sum_{S=0}^{n}|A_{S}|P(\langle t_{i},t_{j}\rangle)\nonumber\\
         &=& \sum_{s\in[0,\delta_{1}]}|A_{S}|P(\langle t_{i},t_{j}\rangle)+\sum_{s\in[1-\delta_{2},1]}|A_{S}|P(\langle 
t_{i},t_{j}\rangle)\nonumber\\
         &\leq&
         E^{2}(N)(1+O(\frac{1}{n}))+E(N)(1+O(n^c))
\end{eqnarray}
\indent Consequently, by the Cauchy inequality, we have
\begin{eqnarray}
Pr(Sat)&\geq&\frac{E^{2}(N)}{E(N^{2})}\geq
\frac{E^{2}(N)}{E^{2}(N)(1+O(\frac{1}{n}))+E(N)(1+O(n^c))}\nonumber\\
&>&1-\delta
\end{eqnarray}
\begin{equation}
E(N)>\frac{1-\delta+O(n^c)}{\delta-O(\frac{1}{n})}
\end{equation}
\indent Putting
$\frac{1-\delta+O(n^c)}{\delta-O(\frac{1}{n})}=\vartheta$, hence
we have
\begin{equation}
\alpha n\ln n+rn\ln n\ln(1-p)>\ln\vartheta
\end{equation}
\begin{equation}
r<\frac{\ln\vartheta-\alpha n\ln n}{n\ln
n\ln(1-p)}=-\frac{\alpha}{\ln(1-p)}+\frac{\ln\vartheta}{n\ln
n\ln(1-p)}
\end{equation}
\indent So we obtain that
\begin{equation}
r<r_{cr}+\frac{\ln\vartheta}{n\ln n\ln(1-p)}
\end{equation}
\indent Thus when $r<r_{cr}+\Theta(\frac{1}{n\ln n})$
, we have the result $Pr(Sat)>1-\delta$. \\
\textbf{Case 2:} $c\geq0$. \\
\indent When $1\leq t\ll n$, by the right side of (4.10), we can
get
\begin{eqnarray}
&&[1-p+p(1-\frac{t}{n})]^{rn\ln n}(n^\alpha-1)^{t}{n\choose
t}\nonumber\\&=& n^{-rkpt+O(\frac{1}{n})}n^{\alpha
t}(1-\frac{1}{n^\alpha})^{t}\sqrt{2\pi n}(\frac{n}{t})^t(1+O(\frac{1}{n}))\nonumber\\
&\leq&\frac{\sqrt{2\pi n}\cdot
n^{(\alpha+1+O(\frac{1}{n})-rkp)t}}{t^t}(1+O(\frac{1}{n}))
\end{eqnarray}
\indent Now when $n$ is sufficiently large, let
$u_{t}=\frac{n^{(\alpha+1-rkp)t}}{t^t}=\frac{n^{ct}}{t^t}$. Then
$u_{t}=e^{ct\ln n-t\ln t}$. If we put $\omega_{t}=ct\ln n-t\ln t$,
we can get $\omega_{t}'=c\ln n-\ln t-1$, then $\omega_{t}'=0$ when
$t=\frac{n^{c}}{e}$. And it is known that $0\leq c<1$ by Lemma
4.1. So $|A_{S}|P(\langle t_{i},t_{j}\rangle)$ has the maximal
value $\sqrt{2\pi n}\cdot e^{\frac{n^{c}}{e}}$ at the point of
$t=\frac{n^{c}}{e}$. So we can have
\begin{equation}
\sum_{s\in[1-\delta_{2},1]}|A_{S}|P(\langle
t_{i},t_{j}\rangle)\leq E(N)\sqrt{2\pi n}\cdot
e^{\frac{n^{c}}{e}}n(1+O(\frac{1}{n}))
\end{equation}
\indent We use the Cauchy inequality\\
\begin{eqnarray}
Pr(Sat)&\geq&\frac{E^{2}(N)}{E(N^2)}\geq\frac{E^{2}(N)}{(E^{2}(N)+E(N)\sqrt{2\pi
n}\cdot
e^{\frac{n^{c}}{e}}n)(1+O(\frac{1}{n}))}\nonumber\\
&>&1-\delta
\end{eqnarray}
\begin{equation}
E(N)>\frac{1-\delta+O(\frac{1}{n})}{\delta-O(\frac{1}{n})}\sqrt{2\pi
n}\cdot e^{\frac{n^{c}}{e}}n
\end{equation}
\indent Let $\frac{\sqrt{2\pi
}(1-\delta+O(\frac{1}{n}))}{\delta-O(\frac{1}{n})}=\lambda$, then
we get
\begin{equation}
\alpha n\ln n+rn\ln n\ln(1-p)>\ln
\lambda+\frac{n^{c}}{e}+\frac{3}{2}\ln n
\end{equation}

\begin{eqnarray}
r&<&\frac{-\alpha n\ln n+\frac{n^{c}}{e}+\frac{3}{2}\ln n+\ln
\lambda}{n\ln n\ln(1-p)}\nonumber\\
&=&-r_{cr}+\frac{1}{en^{1-c}\ln
n\ln(1-p)}+\frac{3}{2n\ln(1-p)}+\frac{\ln\lambda}{n\ln n\ln(1-p)}
\end{eqnarray}

\indent So when $r<r_{cr}+O(\frac{1}{n^{1-c}\ln n})$, we have
$Pr(Sat)>1-\delta$.\\
\indent Combining the above cases, it is proved that the scaling
window of model RB is
\begin{equation}
W(n,\delta)=(r_{cr}-\Theta(\frac{1}{n^{1-\varepsilon}\ln n}),\ \
r_{cr}+\Theta(\frac{1}{n\ln n})),\nonumber
\end{equation}
where $\varepsilon=\frac{c+|c|}{2}$, $c<1$ and it is obvious that
$|r_{cr}+\Theta(\frac{1}{n\ln
n})-(r_{cr}-\Theta(\frac{1}{n^{1-\varepsilon}\ln
n}))|\rightarrow0$ ($n\rightarrow\infty$).
Thus, we finish the proof of Theorem 3.1.\\
\\
\textbf{Remark 4.1} By Lemma 4.1, we claim that $c$ increases with
$p$ and decreases with $\alpha$. Therefore, when $0\leq c<1$, the
convergence rate of $r_{-}(n,\delta)$ approaching $r_{cr}$
decreases with $p$ and increases with $\alpha$.\\
\\
\textbf{Proof of Theorem 3.2} Similarly, we can also use (4.3) to
obtain that
\begin{equation}
\ln(1-p)<-\frac{\alpha}{r}+\frac{\ln \delta}{rn\ln n}
\end{equation}
\begin{eqnarray}
p&>&1-e^{-\frac{\alpha}{r}+\frac{\ln\delta}{rn\ln
n}}\nonumber\\
&=&1-e^{-\frac{\alpha}{r}}+e^{-\frac{\alpha}{r}}(1-e^{\frac{\ln
\delta}{rn\ln n}})\nonumber\\
&=&p_{cr}+e^{-\frac{\alpha}{r}}[1-(1+O(\frac{\ln\delta}{rn\ln
n}))]\nonumber\\
&=&p_{cr}+\Theta(\frac{1}{n\ln n})
\end{eqnarray}
\indent So when $p>p_{cr}+\Theta(\frac{1}{n\ln n})$, we have
$Pr(Sat)<\delta$. \\
\indent Similar to the proof of Theorem 3.1, when $n$ is
sufficiently large, let $c=\alpha+1-rkp_{cr}$. So by Lemma 4.2 we
can also divide $c$ into two cases, that is to say $c<0$ and
$0\leq c<1$. Therefore, we have the followings.\\
\indent By (4.19), we can get
\begin{equation}
\ln(1-p)>\frac{\ln\vartheta-\alpha n\ln n}{rn\ln
n}=-\frac{\alpha}{r}+\frac{\ln\vartheta}{rn\ln n}
\end{equation}
\begin{eqnarray}
 p&<&1-e^{-\frac{\alpha}{r}+\frac{\ln\vartheta}{rn\ln
n}}\nonumber\\
&=&
1-e^{-\frac{\alpha}{r}}+e^{-\frac{\alpha}{r}}(1-e^{\frac{\ln\vartheta}{rn\ln
n}})\nonumber\\
&=& p_{cr}+e^{-\frac{\alpha}{r}}[1-(1+O(\frac{\ln\vartheta}{rn\ln
n}))]\nonumber\\
&=& p_{cr}-\Theta(\frac{1}{rn\ln n})
\end{eqnarray}
\indent By (4.26), we have
\begin{equation}
\alpha n\ln n+rn\ln
n\ln(1-p)>\ln\lambda+\frac{n^c}{e}+\frac{3}{2}\ln n
\end{equation}
\begin{eqnarray}
p&<&1-e^{-\frac{\alpha}{r}+\frac{\frac{n^c}{e}+\frac{3}{2}\ln
n+\ln\lambda}{rn\ln
n}}\nonumber\\&=&1-e^{-\frac{\alpha}{r}}+e^{-\frac{\alpha}{r}}(1-e^{\frac{\frac{n^c}{e}+\frac{3}{2}\ln
n+\ln\lambda}{rn\ln
n}})\nonumber\\&=&p_{cr}+e^{-\frac{\alpha}{r}}[1-(1+O(\frac{1}{n^{1-c}\ln
n}))]\nonumber\\&=&p_{cr}-\Theta(\frac{1}{n^{1-c}\ln n})
\end{eqnarray}
\indent Thus the results are as follows:
$$Pr(Sat)>1-\delta,\ \ when\ \  p<p_{cr}-\Theta(\frac{1}{n^{1-\varepsilon}\ln n}) ;$$
$$Pr(Sat)\ \ <\ \ \delta,\ \ \  when\ \ \ p>p_{cr}+\Theta(\frac{1}{n\ln n}) ,$$
where $\varepsilon=\frac{c+|c|}{2}$, $c<1$ and
$0\leq\varepsilon<1$.\\
\indent Therefore the scaling window of model RB with respect to parameter $p$ is\\
$$W(n,\delta)=(p_{cr}-\Theta(\frac{1}{n^{1-\varepsilon}\ln n}),\ \
p_{cr}+\Theta(\frac{1}{n\ln n}))$$
\\
\textbf{Remark 4.2} Similar to Remark 4.1, by Lemma 4.2, we obtain
that the convergence rate of $p_{-}(n,\delta)$ approaching
 $p_{cr}$ increases with both $r$ and $\alpha$.\\
\\
\indent Note that especially, when $n\rightarrow \infty$, we have \\
$$Pr(Sat)\rightarrow 0, when\ r>r_{cr}\ \ or\ \ p>p_{cr},$$
$$Pr(Sat)\rightarrow 1, when\ r<r_{cr}\ \ or\ \ p<p_{cr}.$$
\indent This is the result of Xu and Li\cite{XWe}.
\section{Conclusions}
\indent In this paper, we obtain the scaling window of model RB
for which the phase transition point is known exactly. As
mentioned before, the scaling window of random 2-SAT has also been
determined. However, this model is easy to solve because 2-SAT is
in P class. Recently, both theoretical\cite{XWm} and experimental
results\cite{XB} suggest that model RB is abundant with hard
instances which are useful both for evaluating the performance of
algorithms and for understanding the nature of hard problems. As
far as we know, this paper is the first study on the scaling
window of such a model with hard instances. We hope that it can
help us to gain a better understanding of the phase transition
phenomenon in NP-complete problems.
\bibliographystyle{amsplain}

\begin{thebibliography}{10}
\bibitem{A}D. Achlioptas and G. Sorkin, Optimal myopic algorithms
for random 3-SAT, \textsl{Proceedings of the 41st Annual IEEE
Symposium on Foundations of Computing} (2000) 590-600.
\bibitem{B}B. Bollob\'{a}s, Christian Borgs, Jennifer Chayes, J.H.
Kim, and D.B. Wilson, The scaling window of the 2-SAT transition,
\textsl{Random Structures and Algorithms}, 201-256(2001).
\bibitem{Che}P. Cheeseman, B. Kanefsky and W. Taylor,  Where the
really hard problems are, \textsl{Proc. 12th Int. Joint Conf. on
Artificial Intelligence}, 331-337(1991).
\bibitem{Chv}V. Chv\'{a}tal and B. Reed, Mick gets some (the odds are on his
side), \textsl{Proc. 33rd Symposium on the foundations of Computer
Science}, 620-627(1992).
\bibitem{Co}S.A. Cook. The complexity of theorem-proving
procedures, \textsl{Proc.3rd ACM Symposium on Theory of
Computing}, 151-158(1971).
\bibitem{F}Y. Fu and P.W. Anderson, Application of satistical
mechanics to NP-complete problems in combinatorial optimisation,
\textsl{J. Phys. A}, 19:1605-1620(1986).
\bibitem{Ge}I.P. Gent, E. MacIntyre, P. Prosser and T. Walsh, The
constrainedness of search, \textsl{In Proceedings of the 13th
National Conference on AI }, pages 315-320. American Association
for Artifical Intelligence, (1997).
\bibitem{Go}A. Goerdt, A threshold for unsatisfiability, \textsl{Journal of Computer and System
Sciences}, 33:469-486(1996).
\bibitem{Ho}T. Hogg, B. A. Huberman, and C. Williams, Eds.,
Frontiers in problem solving: phase transitions and complexity,
\textsl{Artificial Interlligence} 81, 1996.
\bibitem{Ki}L. M. Kirousis, E. Kranakis, D. Krizanc and Y. C.
Stamatiou, Approximating the unsatisfiability threshold of random
formulea, \textsl{Random Structures and algorithms} 12(1998)
253-269.
\bibitem{K}S. Kirkpatrick and B. Selman, Critical behavior in the
satisfiability of random boolean expressions, \textsl{Science},
264:1297-1302(1994).
\bibitem{M}R. Monasson, R. Zecchina, S. Kirkpatrick, B. Selman,
and L. Troyansky, Determining computational complexity from
characteristic phase transitions, \textsl{Nature},
400:133-137(1999).
\bibitem{S}B. Selman, H. Levesque and D. Mitchell, Hard and easy
distributions of SAT problems, \textsl{Proc. 10th Nat. Conf. on
Artificial Intelligence}, 459-465(1992).
\bibitem{V}Y. Verhoeven, Random $2$-SAT and unsatisfiability,
\textsl{Information Processing Letters}, 72:119-123(1999).
\bibitem{XB}K. Xu, F. Boussemart, F. Hemery and C. Lecoutre,
Random Constraint Satisfaction: Easy Generation of Hard
(Satisfiable) Instances, \textsl{Artificial Intelligence},
171(2007):514-534, Earlier version appeared in Proc. of 19th
IJCAI, pp.337-342, Scotland, 2005.
\bibitem{XWe}K. Xu and W. Li, Exact phase transition in random
constraint satisfaction problems, \textsl{Journal of Artificial
Intelligence Research}, 12:93-103(2000).
\bibitem{XWm}K. Xu and W. Li, Many hard
examples in exact phase transitions, \textsl{Theoretical Computer
Science}, 355:291-302(2006).

\end{thebibliography}

\end{document}